\documentclass[epj]{webofc}
\usepackage[varg]{txfonts}   
\wocname{EPJ Web of Conferences}
\woctitle{CONF12}
\begin{document}
\selectlanguage{english}
\title{Effects of Induced Surface Tension in Nuclear and Hadron Matter }
%
%

\author{V.V. Sagun\inst{1,2}\fnsep\thanks{\email{violettasagun@tecnico.ulisboa.pt}} \and
        K.A. Bugaev\inst{1} \and
        A.I. Ivanytskyi\inst{1,3} \and
        D.R. Oliinychenko\inst{1,4} \and
        I.N. Mishustin\inst{2,3}
}

\institute{Bogolyubov Institute for Theoretical Physics, Metrologichna str. 14$^B$, Kyiv 03680, Ukraine
\and
Centro Multidisciplinar de Astrof{\'i}sica, Instituto Superior T{\'e}cnico, Universidade Tecnica de Lisboa,
Av. Rovisco Pais 1, 1049-001 Lisboa, Portugal
\and
Astronomical Observatory of Taras Shevchenko National University of Kyiv, Observatorna 3, Kyiv, 04053, Ukraine
\and
           FIAS, Goethe-University, Ruth-Moufang Str. 1, 60438 Frankfurt upon Main, Germany
\and
           Kurchatov Institute, Russian Research Center, Akademika Kurchatova Sqr., Moscow, 123182, Russia
}

\abstract{%
Short range particle repulsion is rather important property of the hadronic and nuclear matter equations of state. We present a novel equation of state which is based on the virial expansion for the multicomponent mixtures with hard-core repulsion. In addition to the hard-core repulsion taken into account by the proper volumes of particles,  this  equation of state explicitly contains the surface tension which is  induced by another part of the  hard-core repulsion  between particles. At high densities the induced surface tension vanishes and  the excluded volume treatment  of hard-core repulsion is switched to its proper volume treatment. Possible applications of this equation of state to a description of hadronic multiplicities measured in A+A collisions, to an investigation of the nuclear matter phase diagram properties and to the neutron star interior modeling are discussed. 
}
\maketitle
\section{Introduction}
\label{intro}

The excluded volume effects play a significant role in description of the experimental data measured in the  nucleus-nucleus (A+A) collisions, in the studies of the nuclear matter phase diagram and in the modeling of the neutron star interiors. Studies of such systems at high baryonic densities at which the usual Van der Waals 
approximation is inapplicable  require more elaborate equations of state (EoS). In  the vast majority of models, however,  such studies are performed using the Van der Waals approach which, unfortunately,  is inapplicable at the  particle densities that are close to  the transition region to quark gluon plasma (QGP). Foremost, the problem  is related to the wrong values of the third, the fourth and  higher virial coefficients generated by  the Van der Waals EoS. As it was shown in \cite{Bugaev:13NPA} the solution of this problem requires to account for the fact that  at low densities an interparticle hard-core repulsion is well described  by the excluded volume approximation, whereas the high density regime is controlled by the proper volume of particles.  The Van der Waals prescription is not able to switch between these two regimes and, therefore, it requires for an improvement. Another  problem which is typical for all EoS with the hard-core repulsion are their non-causal behavior at high particle densities. 

This very fact motivated us to develop a novel EoS based on the analysis of the virial expansion for the multicomponent mixtures, i.e. for any number of hard-core radii of particle species. We also require that such an EoS  should  be able to reproduce (at least) third and fourth virial coefficients of the gas of hard spheres. Below it is shown that the latter requirement allows us to formulate a thermodynamically consistent EoS 
 which obeys  causality up
to several normal nuclear densities. 
A significant advantage of the present EoS which is important for practical applications  is that it has a form of  two coupled nonlinear equations for any, even infinite,  number of particle species. The contribution of surface tension induced by the particle interaction is a principally new element of  suggested  approach. 

The work  is organized as follows. In the next section the theoretical basis of the present  model is given.
In Sections \ref{sec-2} and  \ref{sec-3} we present the application of the proposed model to the hadron and nuclear matter EoS, respectively. Section \ref{sec-4} is  devoted to the conclusions.

\section{Model formulation}
\label{sec-1}

The present  model is formulated on the basis of the consistent treatment  of the second virial coefficients for an  ensemble of an infinite number of hard-core radii either nuclear or hadron fragments of all sizes. Such a  virial expansion allows us to explicitly  account  for  the many-body  effects  and  to deduce  that the hard-core interaction between  the constituents   induces  an  additional contribution into the surface tension free energy. Thermodynamically consistent  EoS  developed in \cite{Bugaev:13NPA} is a system of coupled  equations between the  pressure  $p$ of considered system  and  the induced surface tension coefficient $\Sigma$  which are as follows
\begin{eqnarray}
\label{EqI}
p &=& T \sum_{k=1}^N \phi_k \exp \left[ \frac{\mu_k}{T} - \frac{4}{3}\pi R_k^3 \frac{p}{T} - 4\pi R_k^2 \frac{\Sigma}{T} \right]
\,, \\
\label{EqII}
\Sigma &=& T \sum_{k=1}^N R_k \phi_k \exp \left[ \frac{\mu_k}{T} - \frac{4}{3}\pi R_k^3 \frac{p}{T} - 4\pi R_k^2 \alpha \frac{\Sigma}{T} \right] \,,
\end{eqnarray}
where $\mu_k$, $m_k$ and $R_k$ are, respectively,  the chemical potential, the mass and the hard-core radius of the $k$-sort of particles. The actual parameterization of the one-particle thermal density $\phi_{k}(T,m,g)$ corresponding to the
particle of the mass $m_{k}$ and the degeneration factor $g_{k}$ depends on the nature of constituents and, hence,  it is discussed below in details.
The summations in  Eqs. (\ref{EqI}) and (\ref{EqII})  are made over all sorts of particles and their antiparticles are considered as independent species. 

The dimensionless  parameter $\alpha$ is  introduced  in (\ref{EqII}) due to the freedom of the Van der Waals extrapolation to high densities \cite{Bugaev:13NPA}.  The  parameter $\alpha$ accounts for the  high density terms  which  modify  the  Van der Waals  EoS  to a more realistic one. As was established in \cite{Bugaev:13NPA} to reproduce the physically correct phase diagram properties of nuclear matter such a parameter should obey the inequality  $ \alpha >1 $. The  physical meaning of $\alpha$ is a switcher between the excluded volume and the proper volume regimes. 
To see this we consider the following relation
\begin{eqnarray}
\label{EqIIIn}
\Sigma & =   &  p \, R  \, \exp \left[ - 4 \pi R^2 \cdot (\alpha-1)\, \frac{\Sigma}{T} \right] \, ,
\end{eqnarray}
between the total pressure $p$ 
and the induced surface tension coefficient $\Sigma$ for   the one component case, i.e. when all particles have  the same  hard-core radius $R$. 
Eq. (\ref{EqIIIn})  immediately  follows from the system (\ref{EqI}-\ref{EqII}) for the same hard-core radius of all particle.  Using the relation (\ref{EqIIIn})  one can 
rewrite the system pressure for one component case as
\begin{eqnarray}
\label{EqIVn}
p & =   & T \sum_{k=1}^N \phi_k  \exp\left[ \frac{{\mu_k}}{T} - v^{eff}  \frac{p}{T}\right]   \,, \quad v^{eff}  ~ =~      v \left[1 +  3  \cdot \exp \left( -3  v  \cdot (\alpha-1)\, \frac{\Sigma}{T\,  R} \right) \right] \,, 
\end{eqnarray}
where we introduced  an effective excluded volume of hadrons $v^{eff}$ which is defined by  their proper volume $v=\frac{4}{3} \pi R^3$.  It is easy to see that in the  low density limit $\mu_k \rightarrow - \infty$ and, hence,  one finds $\frac{\Sigma v}{T  R} \rightarrow 0$  and  $v^{eff}  \simeq 4  v $, i.e. Eq. (\ref{EqIVn}) for $v^{eff} $ correctly reproduces the excluded volume of the one component case.  In the  high density limit
 $\mu_k/T \gg 1$ and, hence,  
$\frac{\Sigma  v}{T  R} \gg 1$, i.e.  for $\alpha >1$ the exponential function on the right hand side of  Eq. (\ref{EqIVn}) vanishes and  
the effective excluded volume becomes equal to the proper volume, i.e. $v^{eff} \simeq  v $. 

\begin{figure}[!ht]
\centerline{
\includegraphics[width=80mm]{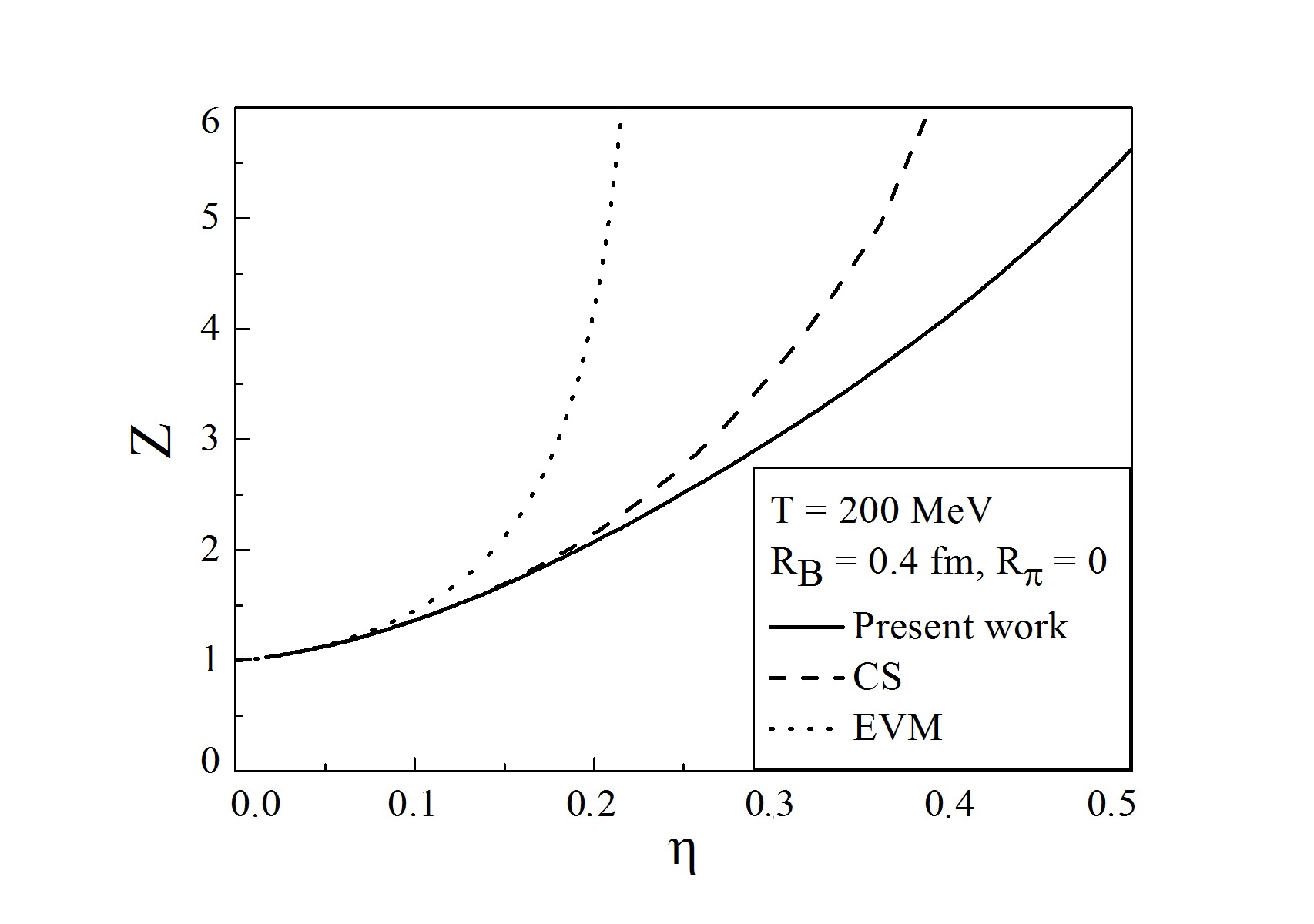}
\includegraphics[width=80mm]{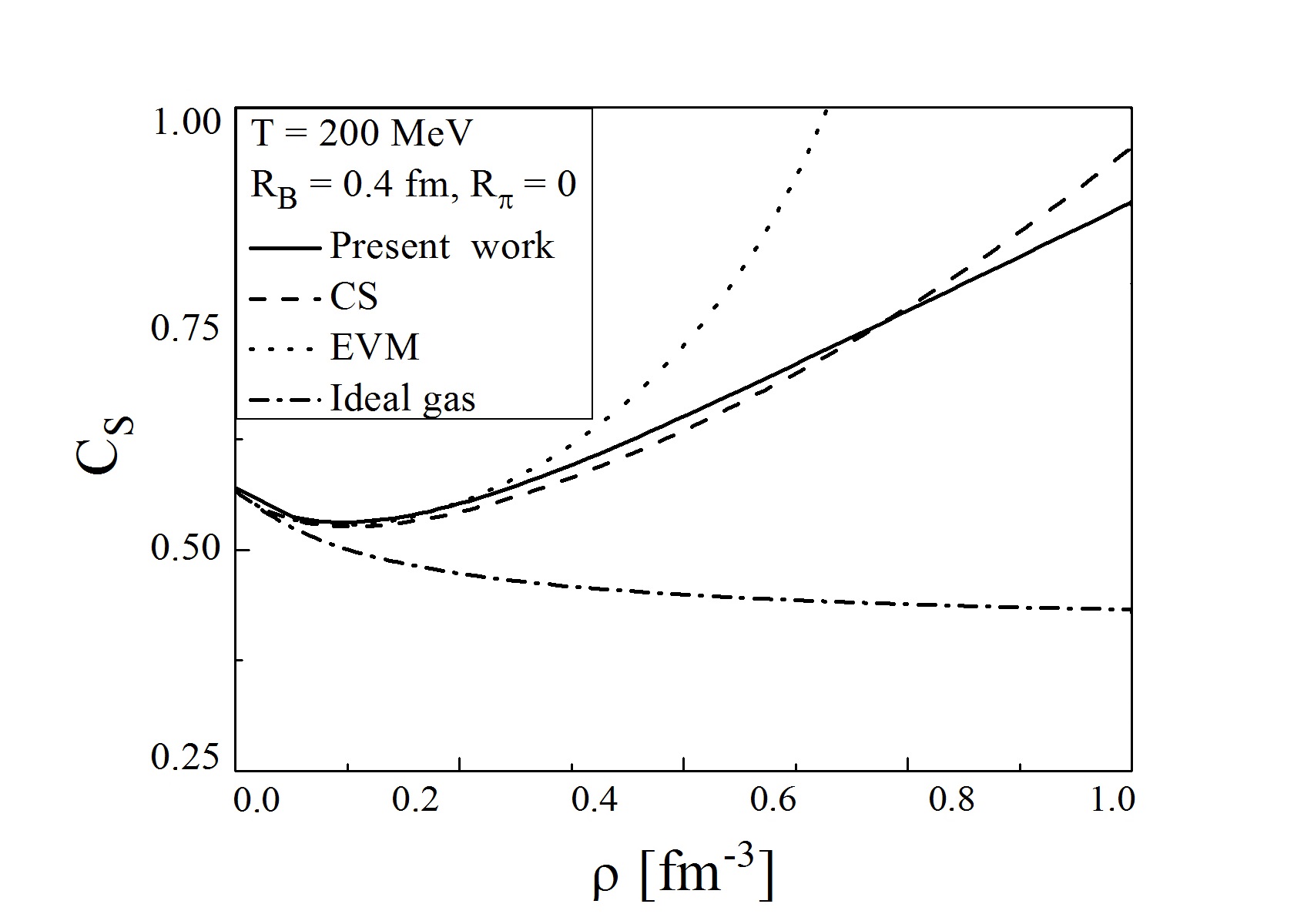}
  }
 \caption{Compressibility factor Z for the mixture of  point-like pions and
nucleons and $\Delta$(1232) baryons having the hard-core radius of 0.4 fm is shown for different EoS
as a function of baryon packing fraction $\eta$ (left panel). The Van der Waals EoS (dotted curve), the IST EoS
(solid curve) and CS EoS (long dashed curve) are shown for T = 200 MeV.
The speed of sound as a function of baryonic density is shown for the same EoS as in the left panel and with the same notations (right panel). The dotted-dashed curve shows the speed of sound for point-like pions
and baryons.
 }
\label{Fig0}
\end{figure}

The value of $\alpha$ was fixed
by  comparing   the   system  (\ref{EqI}-\ref{EqII}) with the induced surface tension (IST EoS hereafter) for the point-like pions and for the nucleons and $\Delta$(1232) baryons having the same hard-core radius 0.4 fm with the famous Carnahan-Starling (CS) EoS \cite{CSeos}.
 As one can see from  Fig. \ref{Fig0} up to the packing fraction of particles $\eta =  v \, \rho \sim$ 0.22 ($v=\frac{4}{3} \pi R^3$ is the proper volume of baryons  and $\rho$ is the baryonic  density) IST EoS for $\alpha=$1.25 reproduces both the compressibility factor $Z$ and the speed of sound $c_S$ of the CS EoS. Such value of $\eta$ corresponds to five values of normal nuclear density.
These figure also shows that the excluded-volume model (EVM), or the usual Van der Waals EoS without attraction,  can be used up to  $\eta \simeq $ 0.11.

In order to reveal the reason for such a good correspondence between the CS EoS
and the IST EoS with $\alpha=$1.25 we calculate the third and fourth virial coefficients of the system
(\ref{EqI}-\ref{EqII}) for the same hard-core radius R and for the same (baryonic) charge of particles
and found 
\begin{eqnarray}
\label{Eq0}
p &=&T \, \rho \, (1+ 4 \, v \, \rho +  B_{3}(\alpha)\, v^{2} \, \rho^{2} + B_{4}(\alpha)\, v^{3} \, \rho^{3} + ...), \ \\
B_{3}(\alpha) &=&  \left[16 - 18 (\alpha-1) \right] v^2 \,, \quad  B_{4}(\alpha) = \left[64 + \frac{243}{2}(\alpha-1)^2 - 216\, (\alpha-1) \right] v^3
\end{eqnarray}
Comparing this result with the virial coefficients $B_3^{hs} = 10 \, v^{2}$ and $B_{4}^{hs} = 18.36 \, v^{3}$  of the one component gas of hard spheres
one finds  
 two solutions $\alpha_{1}\simeq$1.245  and $\alpha_{2}\simeq$2.533 of the equation $B_{4}(\alpha) = 18.36 \, v^{3}$. Since $B_{3}(\alpha=\alpha_{1}) \simeq 11.59 \, v^{2} $ and $B_{3}(\alpha=\alpha_{2})\simeq -11.59 \, v^{3} $, it is evident that the correct root is $\alpha=\alpha_{1}\simeq1.245$. This is an explanation for a very good correspondence between the CS EoS \cite{CSeos} and the IST EoS with $\alpha=$1.25. In fact, $B_{3}(\alpha=1.25)\simeq  11.5 \, v^{2} $ and $B_{4}(\alpha=1.25) \simeq  17.59 \, v^{3}$. In other words, the one component IST EoS with a single additional parameter $\alpha$ is able to 
 simultaneously reproduce the third virial coefficient of the gas of hard spheres with the relative error +15\% and the fourth virial coefficient with the relative error - 4.1\%.

The IST EoS can be  applied to  study  the properties of the hadron resonance gas  and  the ones of  nuclear matter. 
The new EoS  is essentially more effective compared to  the traditional multicomponent Hadron Resonance Gas Model (MHRGM) \cite{PBM06, Horn, SFO, Veta14} and traditional  Statistical Multifragmentation Model (SMM) \cite{Bondorf:95, Sagun:Bugaev07, Sagun:Bugaev00,Sagun:Reuter01},   since  one can easily use  in it an arbitrarily large number of independent hard-core radii.

\section{HRGM with the induced surface tension}
\label{sec-2}

We apply the IST EoS to the description of the hadron multiplicities measured  in the central collisions of  heavy ions. The total   chemical potential of hadron sort $k$ is
\begin{eqnarray}
\label{EqIII}
\mu_k = \mu_B B_k + \mu_{I3} I_{3k} + \mu_S S_k \,,
\end{eqnarray}
where $B_k$, $\mu_B$, $S_k$, $\mu_S$, $I_{3k}$, $\mu_{I3}$ are, respectively, the baryonic, the strange and the  isospin third projection charges and chemical potentials.
The one-particle thermal density $\phi_k$ in Eqs. (\ref{EqI}) and (\ref{EqII})  accounts for  the  Breit-Wigner  mass attenuation and is written in the Boltzmann approximation
\begin{eqnarray}
\label{EqIV}
\phi_k = g_k  \gamma_S^{|s_k|} \int\limits_{M_k^{Th}}^\infty  \,  \frac{ d m}{N_k (M_k^{Th})} 
\frac{1}{(m-m_{k})^{2}+\Gamma^{2}_{k}/4} 
\int \frac{d^3 p}{ (2 \pi)^3 }   \exp \left[ -\frac{ \sqrt{p^2 + m^2} }{T} \right] \,,
\end{eqnarray}
where $g_k$ and $m_{k}$ are, respectively, the  degeneracy factor and the mass of the $k$-sort  of hadrons, $\gamma_S$ is the strangeness suppression factor \cite{Rafelski}, $|s_k|$ is the number of valence  strange quarks and antiquarks  in this kind of hadrons, ${N_k (M_k^{Th})} \equiv \int\limits_{M_k^{Th}}^\infty \frac{d m}{(m-m_{k})^{2}+\Gamma^{2}_{k}/4} $ denotes a corresponding normalization, while $M_k^{Th}$  the decay threshold mass of the hadrons of $k$-sort.

Experimentally detected hadron multiplicity of each hadron  is  the sum of a thermal
component and a component resulting from hadron decays \cite{PBM06}. The
effect of resonance decays $Y\rightarrow X$ to final hadron
multiplicities is taken into account as
follows:%
\begin{equation}
\label{EqX} n^{\rm fin}(X) = \sum_Y BR(Y \to X) n^{th}(Y),
\end{equation}
where $BR(Y\rightarrow X)$ is the probability that hadron~$Y$ decays
into hadron~$X$.\,\,In addition, it is supposed for convenience that
$BR(X\rightarrow X)=1$.\,\,All the parameters used in the fitting of
data (the masses $m_{i}$, the widths $\Gamma_{i}$, the degeneration
factors $g_{i}$ and the probabilities of decays for all strong
decay channels) were taken from the particle tables of the
thermodynamic code \textsc{THERMUS} \cite{THERMUS}.

 The best fit criterion is a minimum  of $\chi^2 = \sum_k  \frac{(r^{theor}_k - r^{exp}_k)^2}{\sigma^2_k} $, where $r_k^{exp}$ is an experimental value of k-th particle ratio, $r_k^{theor}$ is our prediction and $\sigma_k$ is a total error of experimental value.

The proposed IST EoS was used to fit  111   independent  hadronic multiplicity ratios measured  in the  central nuclear collisions  for the center of mass collision energies   $\sqrt{s_{NN}} = $ 2.7, 3.3, 3.8, 4.3, 4.9, 6.3, 7.6, 8.8, 9.2, 12, 17, 62.4, 130 and 200 GeV (for the details of fitting procedure see  \cite{Horn, SFO, Veta14}).  The data sets were taken from Ref. \cite{Veta14}. Then we compared the obtained results with  the ones  found  by the  MHRGM with  the hard-core radii of Ref. \cite{Veta14} (radii from UJP hereafter). 
The best  global fit  of all hadronic multiplicities was found for the following values of 
hard-core radii (new radii hereafter)  of baryons $R_{b}$=0.365 fm, mesons $R_{m}$=0.42 fm, pions $R_{\pi}$=0.15 fm, kaons $R_{K}$=0.395 fm and $\Lambda$-hyperons $R_{\Lambda}$=0.085 fm with the total  $\chi^2/dof=57.099/55 \simeq 1.038$.

\begin{figure}[htbp]
\centerline{
\includegraphics[width=70mm]{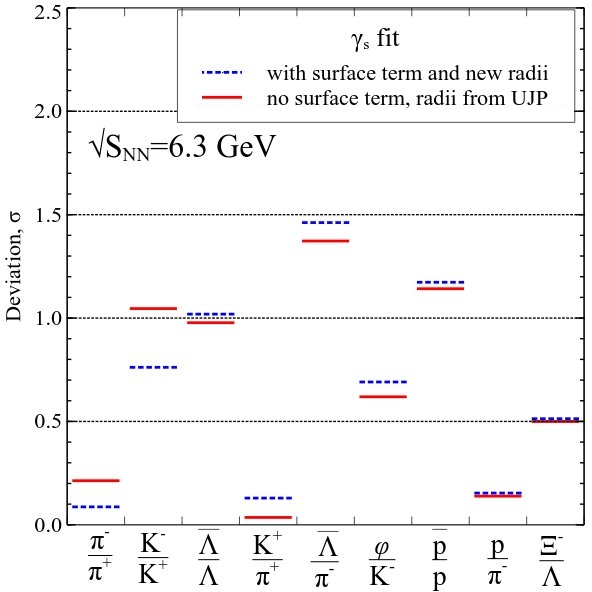}
  \hspace*{0.11cm}
\includegraphics[width=70mm]{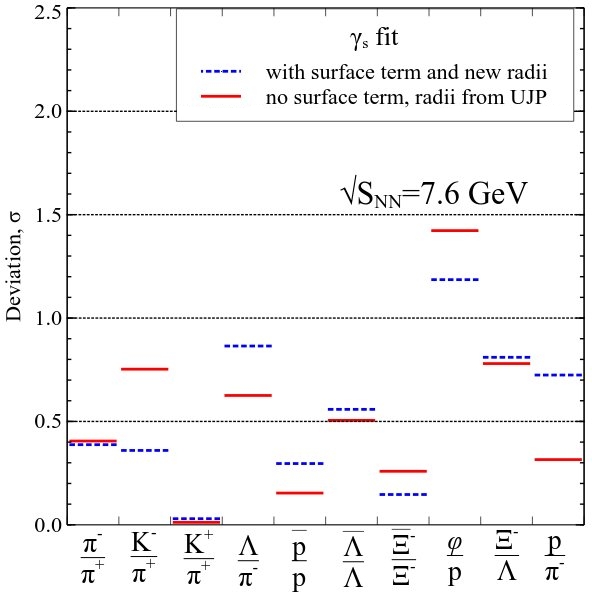}
  }
 \caption{Deviation of theoretically predicted hadronic yield ratios from experimental values in the units of experimental error $\sigma$ for $\sqrt{s_{NN}} = 6.3$ GeV (left panel) and  $\sqrt{s_{NN}} = 7.6$ GeV (right panel). Solid lines correspond to the original MHRGM with $\gamma_{S}$ fit \cite{Veta14}, while the dashed lines correspond to the IST EoS  fit with the $\gamma_{S}$ parameter.
 }
\label{Fig1}
\end{figure}

\begin{figure}[htbp]
\centerline{
\includegraphics[width=70mm]{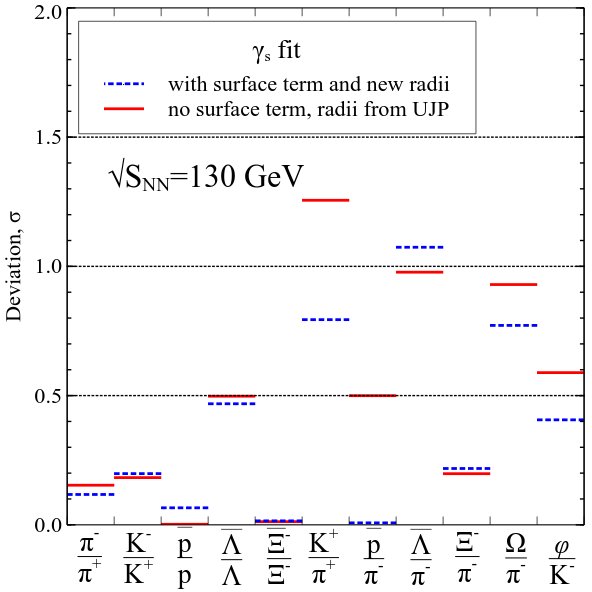}
  \hspace*{0.11cm}
\includegraphics[width=70mm]{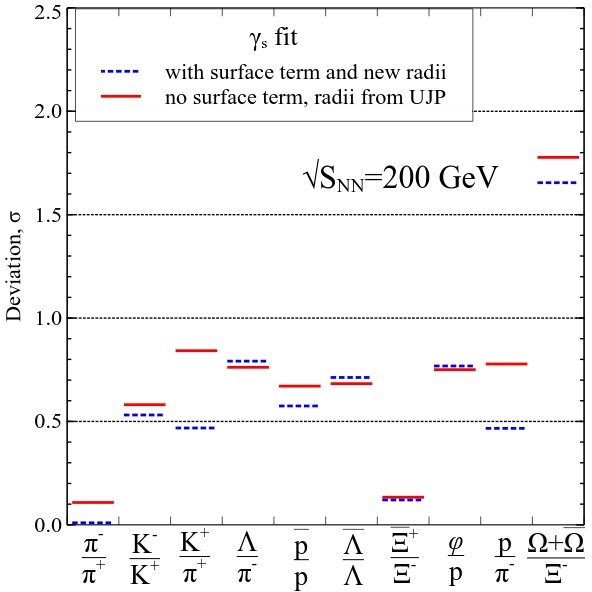}
  }
 \caption{Same as in Fig. \ref{Fig1}, but  for $\sqrt{s_{NN}} = 130$ GeV (left panel) and  $\sqrt{s_{NN}} = 200$ GeV (right panel). 
 }
\label{Fig2}
\end{figure}

 The most prominent examples of the fit results are shown in Figs. \ref{Fig1} and \ref{Fig2} together with the results taken  from Ref. \cite{Veta14}. As one can see from these figures
 some ratios are improved compared to the MHRGM, while other ones  are worsened, but  in general there is no drastic  change 
 of  $\chi^2/dof$.  The most remarkable improvement is gained by the ratio  $\phi / p$ at  at the center of mass collision energy $\sqrt{s_{NN}} = 7.6$ GeV (see the right panel of Fig. \ref{Fig1}) and  the ratio $K^-/\pi^+$ at  $\sqrt{s_{NN}} = 130$ GeV
 (see the left panel of Fig. \ref{Fig2}).  
The  fit results for other collision energies  obtained by the  MHRGM and  by the IST EoS are hardly distinguishable from each other, as it was pointed out above,  the number of equations of the system  (\ref{EqI}-\ref{EqII}) to be solved does not depend 
on the number of different hard-core radii and, therefore,  compared to the  MHRGM with five equations for  five hard-core  radii  the IST EoS is essentially simpler.  To appreciate its advantages below we consider the case of infinite number of hard-core radii.

\section{SMM with the induced surface tension}
\label{sec-3}

The IST EoS  was applied to the description of the nuclear matter properties on the basis of an exactly solvable version of  the   SMM \cite{Sagun:Bugaev00,Sagun:Reuter01}. Such a  model deals with the nucleons with the mass $m \simeq 940$ MeV and the eigen volume $V_{1}=\rho^{-1}_0$ (here $\rho_0 \simeq 0.16 $ fm$^3$ denotes the normal nuclear density  at $T=0$ and zero pressure) and the composite nuclear fragments of any number of nucleons  $k\ge2$. Their  proper  volume is  $V_k=kV_{1}$ and the corresponding surface area is  $S_k$. 

To connect   the above system of equations (\ref{EqI}-\ref{EqII}) for pressure and induced surface tension coefficient with   the gaseous phase  pressure of  the SMM, we used the  parameterization of the one-particle thermal  densities of all  $k$-nucleon fragments as  
\begin{eqnarray}
\label{EqIXa}
 \phi_{1}  &= & z_1 \left[ \frac{m T}{2\pi}\right]^{\frac{3}{2}}  \exp \left[   -  \frac{\sigma (T)}{T}   \right]\, , 
 \\
  \phi_{k \ge 2}  &= & g  \left[ \frac{m T}{2\pi}\right]^{\frac{3}{2}}  \frac{1}{k^\tau} 
  \exp \left[  \frac{\left( k\, p_L V_1 - \mu_k \right)}{T} -  \frac{\sigma (T)}{T}k^{\frac{2}{3}}  \right] \,, 
  \label{EqIX}
\end{eqnarray}
where $z_1 = 4$ is the degeneracy factor of nucleons, while the degeneracy factor for other fragments $g$  is, for simplicity, chosen to be 1 (see a discussion in \cite{Sagun:Bugaev07}). Here $\mu_k$ is the baryonic chemical potential of $k$-nucleon fragment, $\tau \simeq 1.9$ is the Fisher topological exponent and $\sigma (T)$ is the $T$-dependent proper  surface tension coefficient with the following parametrization
\begin{equation}\label{Sagun:IX}
 \sigma (T) =  \sigma_0 \left[ \frac{T_{cep} - T }{T_{cep}} \right]  {\rm sign} ( T_{cep} - T) ~,
\end{equation}
with critical temperature $T_{cep} =18$ MeV and proper  surface tension at zero temperature $\sigma_0 = 18$ MeV. In contrast to the Fisher droplet model \cite{Sagun:Fisher67} and the usual SMM \cite{Bondorf:95}, in the IST SMM the value of the proper surface tension (\ref{Sagun:IX}) is negative above the critical temperature $T_{cep}$. An extended discussion on the validity of such a parameterization can be found in \cite{Sagun:Bugaev13}. 
In order to consider  compressible  nuclear liquid  the following  parameterization of its pressure 
\begin{eqnarray}
\label{Sagun:V}
p_L=\frac{ W(T) +  \mu + a_2 ( \mu -\mu_0)^{2} + a_4 ( \mu -\mu_0)^{4}}{V_1} \,,
\end{eqnarray}
was  suggested  in \cite{Sagun:Bugaev13}.
Here $ W(T) = W_0 + \frac{T^2}{W_0}$ denotes  the usual  temperature dependent  binding energy per nucleon with $W_0 =  16$ MeV \cite{Sagun:Bugaev00} and  the constants  $\mu_0 = - W_0$, $a_2 \simeq 1.233 \cdot 10^{-2}$ MeV$^{-1}$ and $a_4 \simeq 4.099 \cdot 10^{-7}$ MeV$^{-3}$.  These constants  are fixed in order   to reproduce  the properties  of normal nuclear matter, i.e. at vanishing temperature  $T=0$ and normal nuclear density $\rho = \rho_0$ the liquid pressure must be zero.  It is worth to note that such a parametrization of the nuclear liquid pressure describes a compressible nuclear liquid and, in contrast to  the  original SMM  formulation \cite{Bondorf:95}, it   leads to a   nonzero  isothermal compressibility  $K_{T}\equiv\frac{1}{\rho}\frac{d\rho}{dp}\mid_{T}$. 

\begin{figure}[ht]
\centering
\includegraphics[width=71mm]{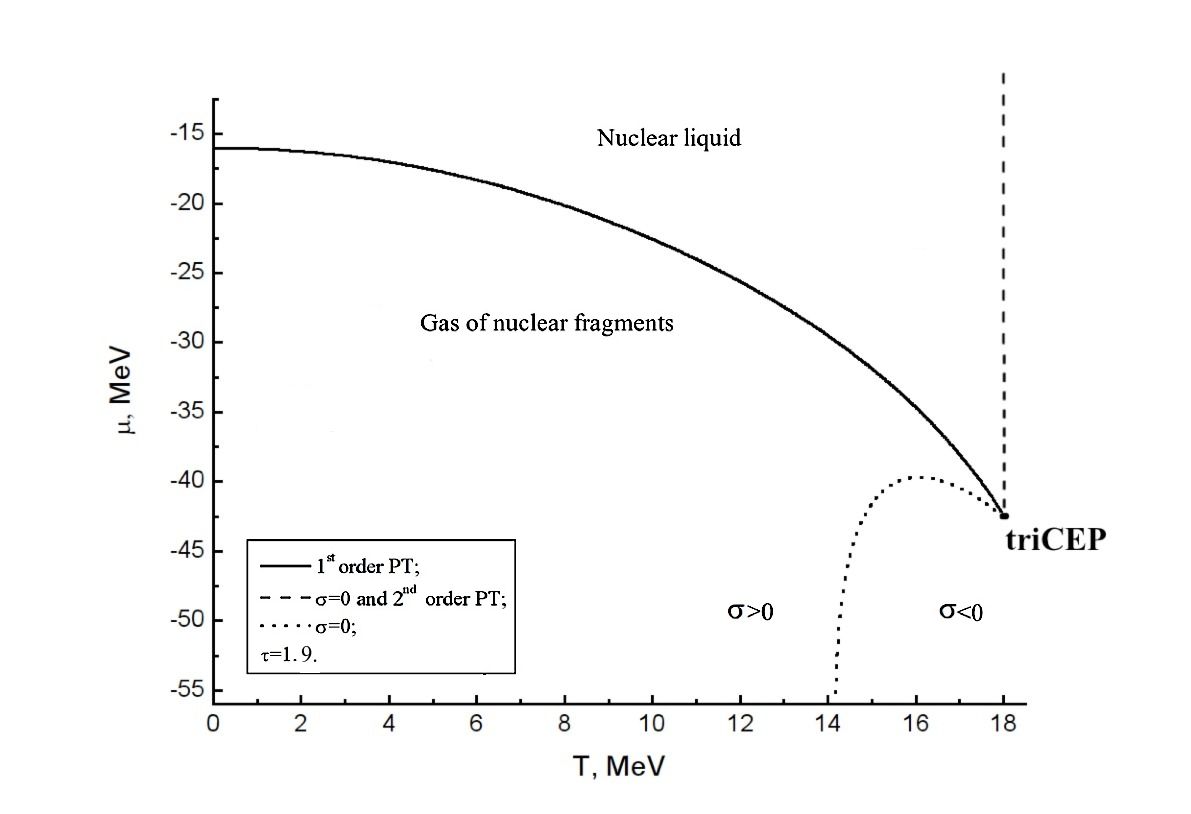}
  \hspace*{-0.35cm}
\includegraphics[width=71mm]{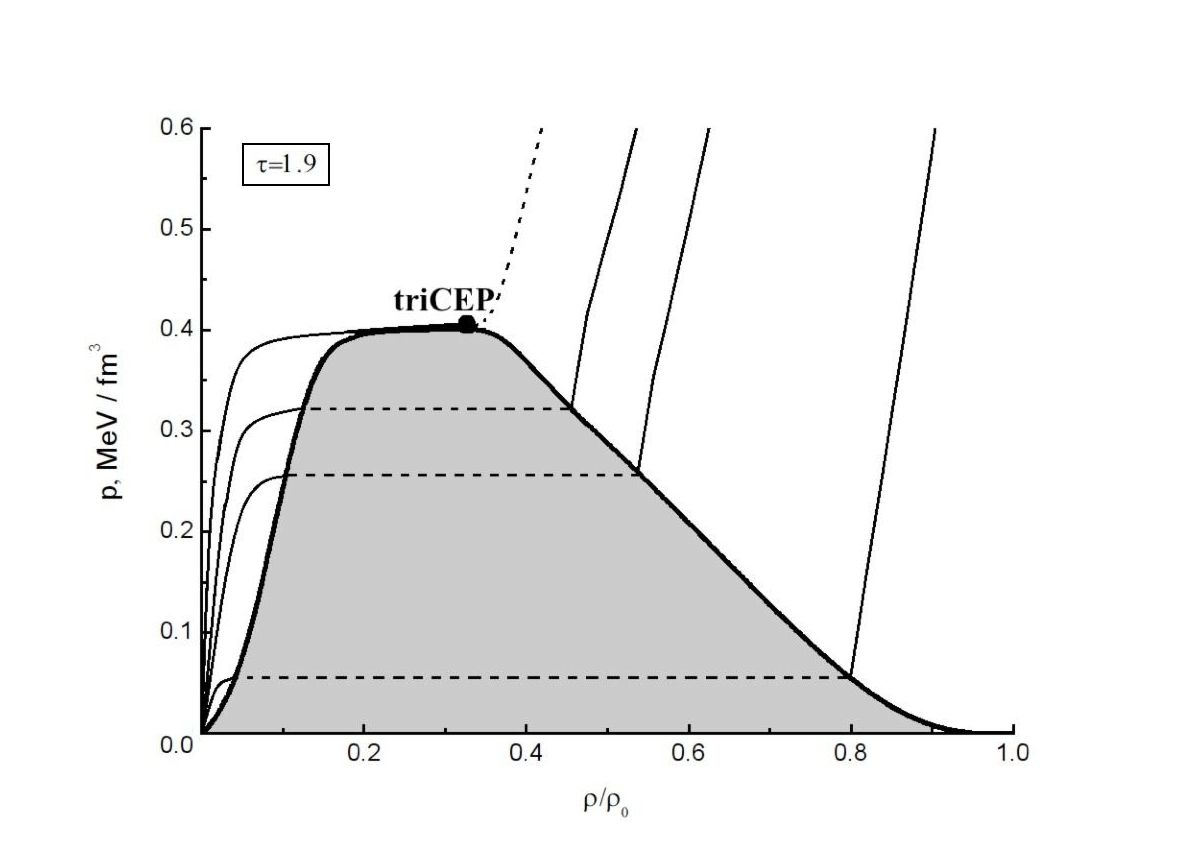}
\caption{Phase diagram in $T-\mu$ plane (left panel) and  $\rho-p$ plane (right panel) for $\nu=2$ and $\tau =1.9$. The  1-st  order PT corresponds to a  solid  curve (left panel) and grey area of a mixed phase  (right panel). The  long dashed line  on the left panel shows  a  2-nd order PT, while the short dashed curve indicates the nil line of the surface tension coefficient.  The isotherms on the right panel are shown for T=11,16,17,18 MeV from bottom to top. While at the critical temperature  $T_{cep}$=18 MeV and density  $\rho/\rho_0  =  1/3 $ there is a triCEP.}
\label{Fig3}
\end{figure}

The IST SMM was solved analytically and the first order phase transition of the liquid-gas type was found in  \cite{Sagun:Bugaev13}. It was proven that such a model has a tricritical endpoint with the temperature $T = 18$ MeV 
and the baryonic density $\rho_{cep} = \rho_0/ 3$. 
The resulting  phase diagram of the IST SMM  in different variables is  shown in Fig.~\ref{Fig3}. 
As one can see, the developed model with the surface tension induced by the repulsive  interaction between the nuclear fragments in  combination with a finite incompressibility of liquid phase has   rather rich phase structure of the nuclear matter phase diagram. 

\section{Conclusions}
\label{sec-4}

In the present work we suggest  thermodynamically consistent approach to account for the  effects of hard-core repulsion 
in the ensemble of constituents (clusters)  of different size. It is important that the IST EoS  allows us to go far beyond the 
usual EVM.
Our analysis shows that interaction between the clusters  leads to  an  additional equation for the induced surface tension coefficient, which at the moment accounts only  for the repulsion between them.
A novel parameter $\alpha$ was introduced  due to the freedom of the Van der Waals extrapolation to high particle densities. On the other hand, this parameter is found to be the ``switcher"  between  the excluded  and proper volume regimes. 
It is shown that a single value of the model parameter $\alpha = 1.25 $ allows us to simultaneously reproduce the third and fourth virial coefficients of the gas of hard spheres with small deviations from their exact values. A  detailed comparison with the famous CS EoS clearly  demonstrates the validity of the IST EoS  at the  packing fractions  0.2-0.22 and its softness  compared to  the  traditional EVM. 
The great advantage of the developed model is that the number of equations to be solved is 2 and it  does not depend on the number of independent hard-core radii.

To employ these advantages in practical applications, the IST EoS was used to study the properties of hadronic and  nuclear  matters. A high quality  description of the experimental hadron multiplicity ratios measured at AGS, SPS and RHIC energies was achieved using the IST EoS  with overall fit quality $\chi^2/dof \simeq 1.04$.

On the basis  of  the  IST EoS  a more realistic version of  the SMM with the compressible nuclear liquid pressure parametrization which generates the tricritical endpoint at the one third of the normal nuclear density  (a typical  value of critical density  for  the liquid-gas phase transitions in the  ordinary liquids) was developed.  It does not lead to an appearance of the non-monotonic isotherms in the mixed phase region which are typical for the mean-field models. This  novel feature makes the present model more realistic than the standard SMM. 

These properties of the proposed EoS make it applicable to the description of the large variety  of physical systems, i.e. to hadron and nuclear  matter phase diagram properties as well the neutron star interior modeling.

\subsection{Acknowledgements}
\label{sec-5}
V.V.S., K.A.B. and  A.I.I.   are   thankful  for the  partial support of the program ``Nuclear matter under extreme conditions''  launched by the Section of Nuclear Physics of National Academy of Sciences of Ukraine.

\end{document}